\documentstyle[preprint,aps]{revtex}

\begin{document}

\title{ Phase Separation Dynamics in a Concentration Gradient}

\author{ A.M. Lacasta$^{1,2}$, J.M. Sancho$^2$ and Chuck Yeung$^{3}$}

\address{ $^1$
Departament de F\'{\i}sica Aplicada,
Universitat Polit\`{e}cnica de Catalunya,\\
Jordi Girona Salgado 31, E-08034 Barcelona, Spain.\\
$^2$
Departament d'Estructura i Constituents de la Mat\`{e}ria, Universitat
de Barcelona,\\
Av.~Diagonal 647,  E-08028 Barcelona, Spain.\\
$^{3}$Department of Physics, University of Toronto,
Toronto, Ontario M5S-1A7 Canada.
}

\date{28 February 1994}

\maketitle

\begin{abstract}

Phase separation dynamics with an initially non-uniform concentration are
studied.  Critical and off-critical behavior is observed simultaneously.
A mechanism for an expanding  phase separated region is demonstrated
and the time dependence of the concentration is determined. The final
equilibrium state consists of a planar interface separating one phase
from the other.  The evolution to this state is characterized by an
experimentally observable flux, $j$, crossing this interface.  We find
that  $j \sim t^{-2/3}$ if patterns are formed in the bulk and $j \sim
t^{-1/2}$ if the bulk remains homogeneous.  The results are explained in
terms of scaling arguments which are confirmed numerically.

\end{abstract}


Phase separation dynamics of a binary mixture is a prototype
nonequilibrium process exhibiting features such as domain
growth, front propagation, pattern formation and pattern selection
\cite{gunton,langer,saarlos}. In a typical experiment, a binary mixture
in the homogeneous phase is rapidly quenched to inside the coexistence
curve.  For a near critical quench convoluted domains of each component
are formed, while in an off-critical quench, droplets of the minority
phase form within the majority phase background.  The phase separation
process exhibits dynamical scaling and the characteristic domain size,
$R$, usually grows as a power law in time.  If the dynamics conserves
the order parameter and hydrodynamics is irrelevant, $R(t)$ grows as
$t^{1/3}$ at late times \cite{lifshitz,oono}.

Usually critical and off-critical quenches occur separately.  However,
if the initial state is inhomogeneous, these processes can be concurrent.
For example, in the binary liquid experiments of Jayalakshmi et al.,
the fluid was prepared with a linear gradient concentration profile
\cite{jayal}.  A spinodal decomposition region was observed in the
middle of the cell with slowly expanding nucleation regions on either
side of the central region \cite{nucleation}.  Kolb et al.\  performed
simulations of the kinetic Ising model using both a linear gradient
profile and a step function profile as the initial states \cite{kolb}.
Their results were interpreted using gradient percolation theory and
they concluded that the concentration profile does not change with time.

In this letter we study phase separation with inhomogeneous initial
conditions via a numerical integration of the time-dependent
Ginzburg-Landau equation for a conserved field (the Cahn-Hilliard
equation) \cite{hohenberg}.  We discuss an additional mechanism for the
expansion, with time, of the area in which phase separation has occurred.
We show that the kinetic Ising model simulations \cite{kolb} describe
early time behavior  and follow the change in the concentration profile
with time.  However, since our method is very effective in obtaining
the long time behavior \cite{oono}, our emphasis, will be on the late
stage dynamics not discussed previously.
To reach the final equilibrium configuration of the two co-existing
phases separated by a planar interface, mass must be transferred from
one side of the system to the other.  We find that the mass
current density, $j$, behaves as $j \sim t^{-2/3}$ if patterns are formed
in the bulk after the quench and $j \sim t^{-1/2}$ if the bulk remains
homogeneous.  The results are explained in terms of scaling arguments
based on, in the first case, the formation of droplets, and, in the
latter case, diffusion.  We confirm the scaling assumptions numerically.

We describe the phase separation dynamics using the
Cahn-Hilliard equation,
\begin{eqnarray}
	\frac{\partial}{\partial t}{c}({\bf r},t)=\nabla^2
	\frac{\delta F(\{c\})}{\delta c({\bf r},t)}
	=  \nabla^{2} \mu( {\bf r}, t ),
		\label{eq:CH}
\end{eqnarray}
where $c({\bf r},t)$ is the deviation of the concentration from its
critical value, $F[c]$ is the Ginzburg-Landau $c^{4}$ free energy.
The local chemical potential, $\mu$, is then
\begin{equation}
	\mu( {\bf r}, t) = \frac{\delta F(\{c\})}{\delta c({\bf r},t)} =
	- c( {\bf r}, t) + c( {\bf r}, t)^{3} - \nabla^{2} c( {\bf r}, t).
\end{equation}
Here $c$, ${\bf r}$ and $t$ are scaled to set all coefficients to
unity \cite{grant}.   We assume a deep quench so that an
additive noise term in Eq.\ (\ref{eq:CH}) can be discarded.  Therefore,
any phase separation must be triggered
by randomness in the initial condition \cite{oono}.

We numerically integrated Eq.\ (\ref{eq:CH})  using an Euler
discretization on an $L_{x} \times L_{y}$ lattice with mesh size $\Delta
x = \Delta y = 1.0$ and time step  $\Delta t = 0.01$.  Periodic boundary
conditions are used in the $x$-direction and reflecting b.c.\ in the
$y$-direction ('cylindrical symmetry') \cite{lacasta}.  Two types of
initial conditions are examined.  We consider a linear gradient profile,
\begin{equation}
	c({\bf r},0) =  c_0 - \frac{2 c_0}{L_{y}} y + \xi( {\bf r} ),
		\label{eq:LINEAR}
\end{equation}
where ${\bf r} \equiv (x,y)$, $c_{0}$ is the absolute maximum of $c$
and $\xi( {\bf r} )$ is an uncorrelated random variable with standard
deviation $\Delta(y)$.  In addition, we also consider a step function
for the initial condition, i.e.,
\begin{eqnarray}
	c({\bf r},0) = c_0 \, \theta (L_{y}/2-y) - c_0 \, \theta (y-L_{y}/2) + \xi(
{\bf r} ),
		\label{eq:STEP}
\end{eqnarray}
where $\theta$ is the step function.  Such an initial condition would be
appropriate if the system is allowed to phase separate and equilibrate
and then quenched to another point inside the coexistence curve.

In Refs.\ \cite{jayal} and \cite{kolb}, it was assumed that the only
effect of the concentration gradient was to change the local average
concentration, $\bar{c}(y)$.  The fact that the region in which phase
separation has occurred grows with time was interpreted to be because the
timescale for phase separation increases with increasing deviation from
the critical concentration \cite{jayal}.  That is, inside the spinodal,
the linear growth timescale diverges as $\bar{c}(y)$ approaches the mean
field spinodal while, outside the spinodal, the nucleation timescale
diverges as $\bar{c}(y)$ approaches the coexistence value.  However,
there is also a second possibility.  Phase separation in the region where
$\bar{c}(y) \approx 0$ (near the critical concentration) may serve as a
seed for phase separation in the rest of the system.  That is, the time
at which phase separation occurs is not determined by the local
$\bar{c}$ but by the fact that phase separation has occurred in an
adjoining region.  To test for this possibility we used a linear gradient
profile (Eq.\ (\ref{eq:LINEAR})) with the initial randomness restricted
to a small strip in the middle of the system.  Specifically, $\Delta(y) =
0.01$ for $ 0.45 L_{y} < y < 0.55 L_{y}$ and $\Delta(y) = 0$ otherwise.
Any phase separation outside the middle strip is therefore initiated
from this strip.

Figure \ref{fig:LINEAR} shows the patterns formed after such a quench
with  $c_{0} = 0.8$.  Immediately after the quench a convoluted structure
similar to that for a critical quench is formed in the central region.
There are well-defined boundaries within which phase separation has
occurred and outside which it has not.  The phase separated region
grows leaving behind a droplet-like structure \cite{rogers}.  This zone
continues to grow until it reaches where the local concentration is of
order the mean field spinodal, $c \approx c_{sp} = \pm 1/\sqrt{3}$.
For intermediate times, a convoluted interface exists in the middle,
separating regions of majority plus phase from that of majority minus
phase.  This interface becomes smoother with time but does not become
planar during our simulations. Even at $t=6400$, it has some curvature
correlated with the remaining droplets.

We repeated the experiment with initial randomness over the whole system.
Very similar behavior was found with the phase separated region expanding
at about the same rate as before.  Therefore we conclude
that, even for uniform randomness, the time at which phase
separation occurs is determined by phase separation in a
neighboring region rather than there being uniform phase separation (based
on the local concentration) over the entire system \cite{jayalcomment}.
Which of these two mechanisms dominate depends on the relative
magnitudes of the initial concentration gradient, the initial randomness
and thermal noise.  Elucidating this dependence should prove interesting
but will not be done here.

Figure \ref{fig:LINEARPROFILE} shows the concentration profile averaged
in the $x$ direction, $\bar{c}(y,t) = L_{x}^{-1} \int^{L_{x}}_{0} dx \;
c( {\bf r}, t )$.  There is no change in the profile at early times.
However, with increasing time, $\bar{c}(y,t)$ approaches its  bulk
equilibrium values of $\pm 1$ first at the boundaries of the system
(where $\bar{c}(y,0)$ is in the nucleation regime).  This equilibrium
region invades the phase separated regions very slowly
since the invasion requires the elimination of droplets.  The Ising
model simulations did not observe this slow change in $\bar{c}(y,t)$
due to the smaller systems and times used \cite{kolb}.

Significantly different behavior was found for the step function initial
condition (Eq.\ (\ref{eq:STEP})).  Figure \ref{fig:STEP} shows the
patterns obtained with $c_{0} = 0.5$ and $\Delta(y) = 0.1$ for all $y$.
Although the central interface is unstable at very early times due to
the flux across it (the Mullins-Sekerka instability) \cite{mullins},
it rapidly becomes planar since the flux decreases rapidly with time.
Droplets then emerge everywhere except in a `depletion' region near the
central interface.  Both the droplet size $R(t)$ and depletion region
thickness, $L_{D}(t)$, grow in time.   A plot of $\bar{c}(y,t)$ (Fig.\
\ref{fig:STEPPROFILE}) also show the substantial difference between
the two initial conditions.  Here $\bar{c}(y,t)$ first approaches the
equilibrium interface profile in the central region. The region in which
$\bar{c}(y,t)$ is $\pm 1$ grows around the central interface and is
precisely the depletion region seen in the patterns. Simulations with
$c_{0} = 0.65 > c_{sp}$ and $\Delta(y) = 0.01$ were also performed.
In this case, no droplets were formed because fluctuations are not strong
enough and one only observes the planar central interface.

Therefore, for $c_{0} < c_{sp}$, one of the dominant late stage processes
is the growth of droplets.  However, since the final equilibrium state is
the two coexisting phases separated by a planar interface, therefore
there must also be a macroscopic flux transferring mass across the
central interface.  (In fact, the total area occupied by the droplets decreases
even though the characteristic size is increasing.)  Therefore a measure
of the approach to the final equilibrium is the total mass transferred
per length, $m(t)$,
\begin{eqnarray}
	m(t) & = & \frac{1}{L_{x}}
	\left( \int^{L_{y}}_{L_{y}/2} \! \! dy
	 -
	\int^{L_{y}/2}_{0} \! \! \! dy  \right)
	\int^{L_{x}}_{0} \! \! \! dx \; [ c( {\bf r}, t ) - c( {\bf r}, 0 ) ],
	\nonumber \\
	& = & 2 \int^{t}_{0} dt' \; j( t' ),
\end{eqnarray}
where $j(  t )$ is the average current density across the middle of the
system,
\begin{equation}
	j( t )
	=  -\frac{1}{L_{x}} \int dx \;
	\left. \frac{\partial}{\partial y} \mu( {\bf r}, t )
	\right|_{y=L_{y}/2}.
\end{equation}
This quantity  can be measured directly.  However, it may be
experimentally simpler to examine a first order transition in which the
motion of the domains are determined by the rate at which heat is removed.
In this case the thermal current is analogous to $j(t)$  discussed above.

Figure \ref{fig:CURRENT} shows $j(t)$ for step function initial conditions
with $c_{0} = 0.5$ and $c_{0} = 0.65$.  For $c_{0} = 0.65$ we find $j(t)
\sim t^{-1/2}$ for all times.  At early times, $j(t)$ for $c_{0} = 0.5$ is
very similar.  However, at $t \approx 100$, droplets begin to appear and
there is a rapid decrease in the current density.  There is a crossover
to the late time behavior, $j(t) \sim t^{-2/3}$, which seems to begin
when the local equilibrium is established.  For finite $L_{y}$, there is
a further  drop in the current density when the last bubble disappears.
This last time depends on $L_{y}$, for $L_{y} = 64$, the final regime
begins at $t \approx 3000$.

To understand the behavior of $j(t)$, we consider the chemical potential
$\mu$ which, in contrast to $c( {\bf r}, t)$, is continuous on all
lengthscales.  If $c_{0}  > c_{sp} + \Delta$ no phase separation occurs
in the bulk phases.  The dynamics are effectively one-dimensional
(see Fig.\ \ref{fig:CURRENT}).  Except at the central interface,
we can expand $\mu$ as $\mu = 2 \delta c + {\cal O}( \delta c^{2} )$
where $c = \pm 1 + \delta c$. (The sign depends on the which side of
the interface one is on.) The change in $c(y,t)$ occurs first near $y=0$
so the appropriate boundary conditions are $\mu(y=\pm \infty, t ) = \pm
\mu_{\infty} \equiv \mp c_{0} ( 1 - c_{0}^{2} )$ while, local equilibrium
gives $\mu(y=0,t) = 0$.  Substituting $\delta c \approx \mu/2$ into the
Cahn-Hilliard equation gives,
\begin{equation}
	\frac{ \partial \mu }{ \partial t }
	= 2 \frac{ \partial^{2} \mu}{ \partial y^{2} }.
\end{equation}
Therefore this regime is diffusion controlled \cite{onuki}.  The asymptotic
current density, $j(t) = -\partial \mu(y,t)/\partial y |_{y=0}$, is
then $j(t) = -\mu_{\infty} /\sqrt{  2 \pi t }$ in agreement with
our numerical results (for $c_{0}$ near 1).

The above analysis no longer holds if patterns are present.  The dynamics
are no longer one-dimensional and the  droplets are important (see Fig.\
\ref{fig:CURRENT}).  Figure \ref{fig:STEPPROFILE} shows that most of
the variation in $\bar{c}(y,t)$ occurs within the depletion width,
$L_{D}(t)$, while in the bulk, the patterns are independent of $y$.
The growth of the droplets in the bulk phases is that of an off-critical
quench \cite{rogers} with $R \sim t^{1/3}$ \cite{lifshitz,oono,rogers}.
Therefore, $\mu_{\infty}$ must be time dependent with local equilibrium
giving $\mu_{\infty} \sim 1/R \sim \mp t^{-1/3}$  \cite{kawasaki}.
To complete the picture we need the behavior of $L_{D}(t)$. In principle,
$L_{D}(t)$ may grow with a dynamic exponent different from $R(t)$.
However, results for the Cahn-Hilliard equation near a wall shows the
exponents are the same in both parallel and perpendicular directions
\cite{ma}.  Therefore we assume that $L_{D}(t) \sim R(t) \sim t^{1/3}$
and the asymptotic current density becomes
\begin{equation}
	j(t) \sim \frac{|\mu|}{ L_{D}(t) } \sim t^{-2/3},
\end{equation}
in agreement with our numerical results.

If our assumptions hold, the chemical potential averaged over
$x$ has the scaling form
\begin{equation}
	\mu(y,t) = t^{-1/3} g( y/ t^{1/3} ).
\end{equation}
Figure \ref{fig:MU} shows the scaling function, $g(z)$, for times
from $t=400$ to $t=6400$.  In agreement with our discussion, most
of the variation in $\mu(y,t)$ is in the depletion
region and approaches time-dependent bulk values outside it.
The collapse of $g(z)$ is reasonable indicating that the scaling
form of $\mu(y,t)$ is correct and that our arguments for
$j(t) \sim t^{-2/3}$  are valid.

In summary, we examine phase ordering dynamics for constant gradient and
step function initial conditions.  We demonstrate an additional mechanism
for the expanding phase separated region observed in experiments.
We discuss how the concentration profile changes with time.  We measure
the flux characterizing the evolution to the final equilibrium state.
Its asymptotic behavior is obtained and understood in terms of scaling
arguments.
\ \\


\noindent
We are grateful to M.\ Rao for useful discussions and constructive
criticisms. We thank R.C.\ Desai and O.\ Sch\"{o}nborn
for careful reading of the manuscript.  We acknowledge the
Direcci\'{o}n General de Investigaci\'{o}n Cient\'{\i}fica y T\'{e}cnica
(Spain) for financial support (Proj.\ No.\ PB90-0030) and the Centre de
Supercomputaci\'{o} de Catalunya (CESCA) for computing support.  C.Y.\
acknowledges the support of Natural Sciences and Engineering Council
of Canada.


\begin{figure}

\caption{
Concentration patterns for a constant gradient initial condition with
$c_0=0.8$, $L_{x} = 256$, $L_{y} = 512$ and $\Delta(y) = 0.01$ for $0.45
L_{y} < y < 0.55 L_{y}$.
\label{fig:LINEAR}
}

\end{figure}

\begin{figure}

\caption{
The concentration profiles for parameters  in Fig.\
\protect\ref{fig:LINEAR} (averaged over five configurations).
The equilibrium values $\bar{c}(y,t) = \pm 1$ are first reached at
the boundaries.  The oscillations at late times are because the bubbles
arrange in rows next to the line where $\bar{c}(y,0) = c_{sp}$.
\label{fig:LINEARPROFILE}
}

\end{figure}

\begin{figure}

\caption{
Concentration patterns for a step function initial condition with
$c_{0}=0.5$, $\Delta(y) = 0.1$  and $L_{x} = L_{y} = 512$.
\label{fig:STEP}
}

\end{figure}

\begin{figure}

\caption{
The concentration profiles for the parameters in Fig.\
\protect\ref{fig:STEP} (averaged over 5 configurations).
\label{fig:STEPPROFILE}
}

\end{figure}

\begin{figure}

\caption{
$j(t)$ vs.\ $t$ in one and two dimensions for step function initial
conditions.  The solid line is $j(t) \sim t^{-2/3}$.  For $c_{0} = 0.65$,
$j(t) \sim t^{-1/2}$.
\label{fig:CURRENT}
}

\end{figure}

\begin{figure}

\caption{
Scaled chemical potential, $g(z) = t^{1/3} \mu( z t^{1/3}, t)$, for the
parameters in Fig.\ \protect\ref{fig:STEPPROFILE}.  The unscaled $\mu$
is shown in the inset.
\label{fig:MU}
}

\end{figure}


\begin{thebibliography}{99}
\bibitem{nucleation}
Nucleation differs from spinodal
decomposition in that an activation barrier is required in nucleation.
\bibitem{jayalcomment}
The expanding phase separated region in the binary liquid experiments
\cite{jayal} were in the nucleation regime and at fairly high
temperatures.  The mechanisms in that experiment may be
different from that discussed here.
\end{thebibliography}
\end{document}